\documentclass[12pt]{article}

\usepackage[utf8]{inputenc}
\usepackage{tabularx,booktabs,colortbl}
\usepackage[frozencache, cachedir=.]{minted}
\renewcommand\listingscaption{Figure}
\usepackage[dvipsnames]{xcolor}
\usepackage{color}
\usepackage{soul}
\definecolor{LightGray}{gray}{0.9}
\usepackage{tikz}
\usetikzlibrary{tikzmark}

\usepackage{amsmath,array,graphicx}
\usepackage{textcomp}
\usepackage{gensymb}
\usepackage{float}
\usepackage{graphicx}
\usepackage{textgreek}
\usepackage{multicol, multirow}
\usepackage[many]{tcolorbox}
\usepackage{makecell}
\usepackage{hyperref}

\usepackage[labelfont=bf]{caption}
\usepackage[textfont=it]{caption}
\usepackage[margin=0.75in]{geometry}
\usepackage[superscript]{cite}
\usepackage{tabularx}
\usepackage{subcaption}
\DeclareCaptionFormat{suboverlay}{\gdef\subcapoverlay{(\thesubfigure) #3\par}}
\DeclareCaptionStyle{suboverlay}{format=suboverlay}

\newcommand{\subcaptionOverlay}[1]{%
  \subcaption{}%
  \begin{tikzpicture}
    \node [inner sep=0,anchor=north west]at (-1.5ex,-2ex) (image) {#1};
    \draw node [black] {\subcapoverlay};
  \end{tikzpicture}%
}

\title{Accelerating active learning materials discovery with FAIR data and workflows: a case study for alloy melting temperatures}

\author{Mohnish Harwani$^{a}$, Juan C. Verduzco$^{b}$, Brian H. Lee$^{b}$, and Alejandro Strachan$^{b}$}

\date{
$^a$ Department of Computer Science \\
$^b$ School of Materials Engineering \\
Purdue University, West Lafayette, Indiana, 47907 USA
\\
}

\begin{document}

\maketitle

\begin{abstract}

Active learning (AL) is a powerful sequential optimization approach that has shown great promise in the discovery of new materials. However, a major challenge remains the acquisition of the initial data and the development of workflows to generate new data at each iteration. In this study, we demonstrate a significant speedup in an optimization task by reusing a published simulation workflow available for online simulations and its associated data repository, where the results of each workflow run are automatically stored. Both the workflow and its data follow FAIR (findable, accessible, interoperable, and reusable) principles using nanoHUB’s infrastructure. The workflow employs molecular dynamics to calculate the melting temperature of multi-principal component alloys. We leveraged all prior data not only to develop an accurate machine learning model to start the sequential optimization but also to optimize the simulation parameters and accelerate convergence. Prior work showed that finding the alloy composition with the highest melting temperature required testing 15 alloy compositions, and establishing the melting temperature for each composition took, on average, 4 simulations. By developing a workflow that utilizes the FAIR data in the nanoHUB database, we reduced the number of simulations per composition to one and found the alloy with the lowest melting temperature testing only three compositions. This second optimization, therefore, shows a speedup of 10x as compared to models that do not access the FAIR databases.

\end{abstract}

\section{Introduction}
\label{introduction_section}

The principles of Findable, Accessible, Interoperable, and Reusable (FAIR) data \cite{wilkinson2016fair} are critical to accelerate progress not only in materials science \cite{brinson2024community, scheffler2022fair} but also across other physical sciences and engineering disciplines. \cite{sinaci2020raw, wise2019implementation, tanhua2019ocean}
Significant progress has been achieved through the development of specialized data platforms \cite{jain2013commentary, saal2013materials, blaiszik2016materials, draxl2018nomad, hunt2022sim2ls} and additional nation-level facilities are called for \cite{national2021materials}.
While the economic benefits of FAIR data have been estimated \cite{Euro_FAIR_Report}, concrete examples showcasing its impact on research and discovery remain notably sparse.
Furthermore, despite extensive efforts dedicated to refining optimization methods, the essential role of data sharing and the necessary infrastructure has often been overlooked. 
The vast majority of research data collected remains private, with the published subset typically embedded in formats like graphs, tables, or text that are not machine-actionable. These practices hamper innovation and leads to considerable duplication of efforts. \cite{brinson2024community}

In this context, we illustrate how the reuse of a FAIR workflow and its associated datasets can dramatically expedite materials discovery tasks, demonstrating a 10-fold increase in efficiency.
Building upon prior datasets that adhere to FAIR principles provides a robust foundation for advanced computational techniques in materials discovery.
The discovery process typically involves an optimization in a high-dimensional space, characterized by variables such as composition, processing conditions, and microstructure.
Relying solely on traditional Edisonian approaches proves impractical and inefficient given the vast array of potential combinations.
To address this challenge, a form of sequential optimization, active learning (AL), has emerged as a powerful tool.
AL require high-quality data to train models that are then used to select the most meaningful experiments next.
AL has been successfully applied in numerous materials science problems, significantly reducing the experiments and resources needed when compared to traditional methods.\cite{settles2009active,kusne2020fly, mcdannald2022fly, moon2024active, yoo2021neural, zhang2019active, kulichenko2023uncertainty, farache2022active, verduzco2021active}. The lack of FAIR data hinders the unimpeded application of AL to a wide range of problems.


Recent applications of AL paired with experimental setups in materials science have demonstrated its effectiveness in the discovery and optimization of new materials.
For instance, Kusne et al. presented an autonomous materials research lab powered by a closed-loop autonomous system for materials exploration and optimization (CAMEO).\cite{kusne2020fly} The CAMEO system was integrated into the synchotron beamline at the SLAC National Accelerator Laboratory and was tasked with finding the composition with the largest optical bandgap in the Ge-Sb-Te ternary, with a new optimal material composition, GST-467, discovered in only 19 iterations. 
Similarly, McDannald et al. developed an autonomous neutron diffraction explorer (ANDiE) \cite{mcdannald2022fly}, a system that allows measurements of materials magnetic transition behaviors to be guided by AL and probabilistic modeling. ANDiE showed a 5-fold reduction in the number of measurements required to find the transition temperature.
Finally, Moon et al. demonstrated an application of AL to search novel perovskite oxides for oxygen evolution electrocatalysis. \cite{moon2024active} Their workflow successfully identified an optimal composition that exhibits superior electrocatalytic performance with the lowest known intrinsic overpotential. 


AL has also been used to enhance physics-based simulations, in particular in the development of interatomic potentials.
Yoo et al. showed the importance of judiciously selecting atomic structures to augment the training set when
developing interatomic potentials. \cite{yoo2021neural}. Similarly, Zhang et al. \cite{zhang2019active} developed the Deep Potential Generator (DP-GEN) 
that actively selects data points that improve model performance, optimizing for accuracy with a minimal amount of reference data. 
Kulichenko et al. \cite{kulichenko2023uncertainty} also used an AL approach that modifies the potential energy surface to favor regions with configurations for which the model is uncertain, such as those associated with rare events. 


Finally, AL has found one of its most important application with the screening of potential candidates to reduce candidate experiments or simulations.
A study by Farache et. al \cite{farache2022active} proposed an AL workflow to identify alloys with high melting temperatures from a predefined set of alloys using molecular dynamics (MD) simulations. Using a random forest model to predict the alloys' melting temperature and corresponding uncertainty, an AL process selects compositions to label (i.e. characterize via MD) through information acquisition functions. These alloys are then simulated, and the predictive model is retrained on this expanded set of `known' alloys, now containing an additional alloy. The model then iterates by making an updated prediction over the `unknown' alloys. Importantly, the AL loop launches an end-to-end, fully autonomous MD workflow to characterize the melting temperature of the desired alloy. This is implemented and published for online simulation \cite{nh_meltheas} as a nanoHUB Sim2L.\cite{hunt2022sim2ls} Importantly, Sim2Ls automatically index and store all input-output pairs of the workflows into a database that also follows FAIR principles. The approach required testing only $\sim$15 compositions to find the highest melting alloy out of 555 possible cases.

This paper extends prior work to identify alloys with high melting temperatures by 1) learning from the prior FAIR data to reduce the number of iterations required find a converged melting temperature value using MD, and 2) improving the AL cycle by starting from all available data from the prior work. We demonstrate that this approach not only significantly decreases the time needed to identify alloys with optimal melting temperatures ($\sim$10x improvement over previous work \cite{farache2022active}), but it also allows for data reuse in subsequent optimizations to find alloys with the lowest possible melting temperatures. The fundamental goal is to demonstrate the reusability of data across different material characteristic optimizations. A single tool can learn from different optimizations across one criteria to create a large queryable database, and use AL to navigate that database on a different criteria, finding optimal alloys in fewer iterations. Throughout this work, we will reference the work by Farache et al. \cite{farache2022active} as \textit{Work 1}.

\section{Methods}
\label{methods_section}

\subsection{FAIR workflows and data: Sim2Ls and ResultsDB}
\label{sim2l_section}

Establishing robust data-sharing practices and employing advanced cyberinfrastructure to support them is essential to enhance research efficiency and foster greater innovation across disciplines. This work leverages nanoHUB's infrastructure for FAIR workflows, simulations, and data. \cite{hunt2022sim2ls} These resources enable best data practices with minimal effort with the goal of creating interconnected research environment where information is readily available and usable by researchers worldwide. Recent advances in the nanoHUB environment make scientific workflows and data FAIR. These technologies, Sim2ls and ResultsDB, allow developers to share their research workflows and contribute to a database of open community-generated data.

Sim2Ls are end-to-end research workflows with formally declared inputs and outputs and are implemented as flexible Jupyter notebooks. Inputs and outputs have well-defined types and metadata. Research workflows can contain data processing steps, analysis, and/or computational simulations. Additionally, Sim2Ls simplify the process of setting up a containerized environment, guaranteeing workflow reproducibility across various systems.Each tool results in an independent dataset with a digital object identifier (DOI) and metadata. Each record includes a serialized query identifier (SQUID), its creation date, and all inputs and outputs. This database can be accessed through a web interface or via a REST API. This approach not only enhances reliability but also fosters transparency in research methodologies. Results of successful workflow execution runs are automatically validated, indexed, and stored in the ResultsDB, an accessible database. 

\subsection{Molecular dynamics}
\label{md_section}

This work builds on data from the \textit{meltheas} Sim2l \cite{nh_meltheas} described in detail in \textit{Work 1}. This Sim2l calculates a melting temperature for multi-principal component alloys (MPCAs) containing Cr, Co, Cu, Fe, and Ni through a solid-liquid coexistence method using molecular dynamics simulations \cite{morris1994melting}. This method, seeks to find the temperature at which solid and liquid samples coexist in equilibrium. Atomic interactions are modeled with a many-body, embedded-atom-method (EAM) potential developed by Farkas and Caro \cite{Brian_farkas}, optimized for elastic properties of near-equiatomic MPCAs. The potential was not optimized for melting temperature calculations, and it tends to overestimate the melting temperature of ternary and quaternary alloys. As was done in \textit{Work 1}, and to restrict the optimization to FCC MPCAs, the design space is limited to alloys where no element exceeds 50 at. \%. We stress that our goal is the demonstrate the effect of FAIR data on AL optimization loops and not to recommend alloys for experimental testing. Systems are generated from the primitive FCC unit cell (lattice parameter = 3.56 Å) replicated 8 × 8 × 18 to create a periodic simulation cell. The Z-direction is normal to the solid–liquid interface. Atom types are assigned randomly following the desired alloy composition.

To create an initial two-phase sample, each half of the system gets assigned a different temperature, with a solid region at temperature $T_{sol}$ and the remainder region at temperature $T_{liq}$ under isobaric conditions using two thermostats. These temperatures are designed to cause half of the sample to melt and after a short equilibration (10 ps long) the system is allowed to reach equilibrium using MD simulations in the isoenthalpic, isobaric (NPH) ensemble. The melting temperature corresponds to the system temperature if, after equilibration, the system include both solid and liquid in steady state. This approach is advantageous compared to other simulation methods such as direct melting as it avoids hysteretic effects. \cite{morris1994melting, morris2002melting} A successful determination of melting temperature depends on the  choice of $T_{sol}$ and $T_{liq}$ as well as the length of the simulation.

\subsection{Data collection and exploration}
\label{datacollection_section}

In \textit{Work 1}, authors trained a RF model using a dataset of 39 alloy compositions to predict the melting temperature of multi-principal component alloys (MPCAs) containing Cr, Co, Cu, Fe, and Ni and efficiently explore the design space using AL. As an initial estimate for the ($T_{sol}$) and $T_{liq}$ temperatures, they used a rule-of-mixture calculation to estimate the melting temperature of the alloy in question and assigned $T_{sol}$ 25\% below this value and $T_{liq}$ 50\% above. These values were adjusted if liquid-solid coexistence was not achieved, the temperatures were increased or decreased by 5\% depending on the final phase obtained and the simulation re-run until an acceptable result was obtained.
In average, obtaining a converged value of the melting temperature required 4 simulations per composition. Across the different simulation times and acquisition functions explored in \textit{Work 1}, a total of 265 different compositions were characterized, requiring 3039 simulations. Their design space consisted of 555 compositions. 

We used the results automatically indexed in nanoHUB's ResultsDB from the \textit{MeltHEAs} Sim2L to improve on the simulations parameters and find the alloy composition with the minimum melting temperature in the fewest possible iterations. Since the nanoHUB Sim2l \textit{MeltHEAs} is open access, we took careful consideration to not include any compositions explored by the community after the original article publication date.

\subsection*{Input temperature modeling}
\label{inputtempmodel_section}

\begin{figure}[H]
\centering
\begin{subfigure}{.45\textwidth}
  \centering
  \subcaptionOverlay{\includegraphics[width=\textwidth]{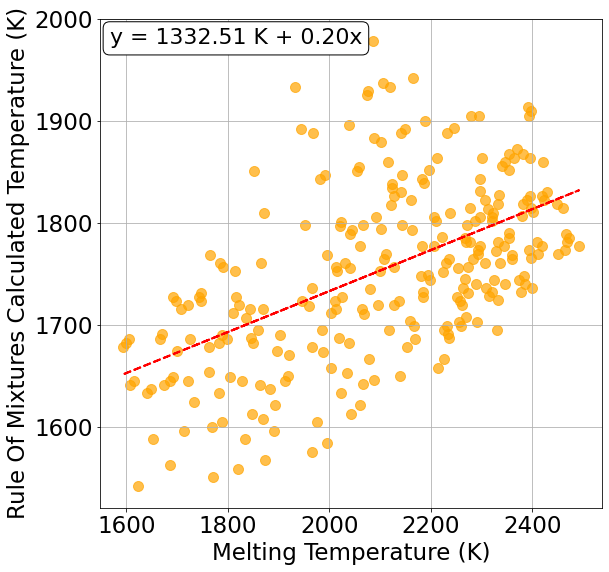}}
\end{subfigure}%
\begin{subfigure}{.45\textwidth}
  \centering
  \subcaptionOverlay{\includegraphics[width=\textwidth]{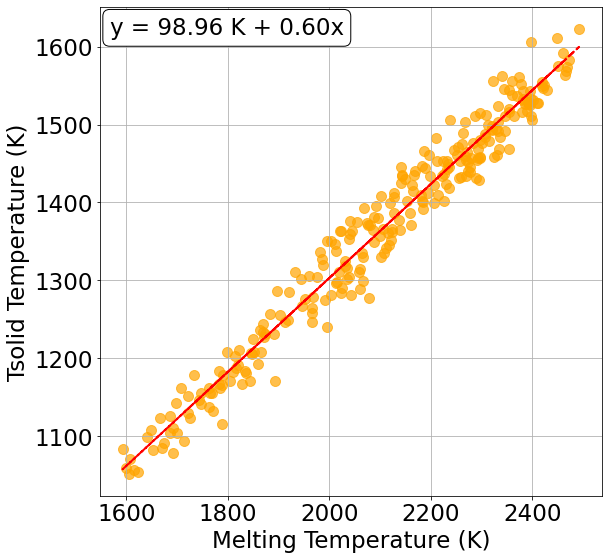}}
\end{subfigure}%
\vspace{0.5em}
\begin{subfigure}{.47\textwidth}
  \centering
  \subcaptionOverlay{\includegraphics[width=\textwidth]{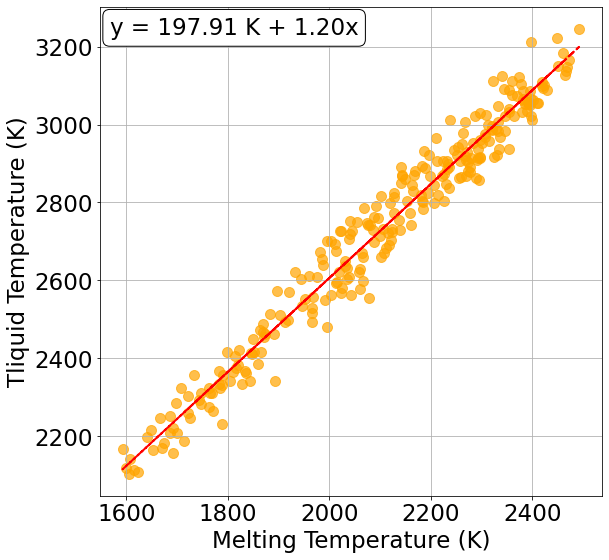}}
\end{subfigure}
\caption{Coexistence method melting temperature against (a) compositionally derived $T_{ROM}$ ($R^2 = 0.2814)$, (b) input temperature $T_{sol}$ ($R^2$ = 0.9884), and (c) input temperature $T_{liq}$ ($R^2$ = 0.9884).}
\label{parity_against_tmelt}
\end{figure}

Our first goal is the develop good estimators for $T_{sol}$ and $T_{liq}$. Figure \ref{parity_against_tmelt}(a) shows that, not surprisingly, that rules of mixtures estimate for the melting temperature used by Farache et al. ($T_{ROM}$) is not very accurate. Figure \ref{parity_against_tmelt}(b-c) show $T_{sol}$ and $T_{liq}$ that resulted in converged MD results as a function of the MD-obtained melting temperature. Clearly, the data in the ResultsDB indicate that $T_{sol}$ and $T_{liq}$ that lead to converged results can be obtained as a linear function of the alloy melting temperature. 

\begin{figure*}[ht]
    \centering
    \includegraphics[scale = 0.45]{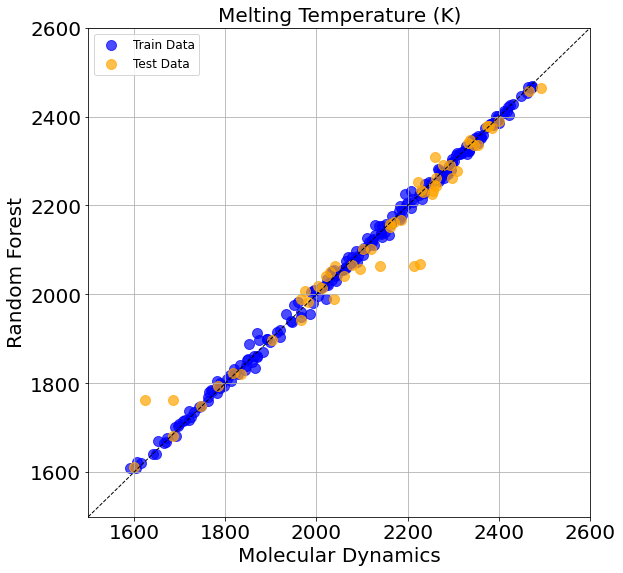}
    \caption{Parity plot between Random Forest predictions of melting temperature and MD simulated melting temperature}
    \label{rf_md_parity}
\end{figure*}

Based on these observations, we hypothesized that accurate starting temperature values could be obtained from a predictive model of $T_{melt}$ (alloy melting temperature) using simple linear models. Based on the approach outlined in \textit{Work 1}, we used an 80/20 train/test split of their data to train a RF model to predict alloy melting temperatures. We kept their exact set of descriptors based on alloy composition (at. \% of each element) and composition-derived features as inputs. This split resulted in a test set of 54 compositions. Our RF model included 100 estimators, which were allowed to grow at to their maximum depth to reduce bias. starting temperatures ($T_{sol}$ and $T_{liq}$) were obtained from the RF ($T_{melt}^{RF}$) prediction using linear models derived from the training data. The predicted melting temperatures are shown in Figure \ref{rf_md_parity}. 

\newpage

\section{Results}
\label{results_section}

\subsection{Number of simulations to obtain a converged melting temperature for a given composition}
\label{subsection_numsim}

We first evaluate our approach to estimate the $T_{sol}$ and $T_{liq}$ inputs to the MD simulations in a subset of previously explored alloys. Authors in \textit{Work 1} estimated the numbers from the RF prediction of the melting temperature ($T_{melt}^{RF}$) as $T_{liq}=1.5 T_{melt}^{RF}$ and $T_{sol} = 0.75 T_{melt}^{RF}$. If the simulation did not result in coexistence and was either too cold or too hot, $T_{liq}$ and $T_{sol}$ were adjusted by increasing or decreasing $T_{melt}^{RF}$ by 5\%. Adjusting $T_{liq}$ and $T_{sol}$ often resulted in overcorrection, thus, in addition to the improved estimates, our approach uses a smaller adjustment of 2.5\%, and, in the case of simulations that achieved coexistence (both fractions of solid and liquid alloy are between 35\% and 65\% of the system) but not a steady state temperature measurement, the simulation time was adjusted by adding 50 ps (the default simulation time is 100 ps). 

\begin{figure*}[ht]
    \centering
    \includegraphics[scale = 0.5]{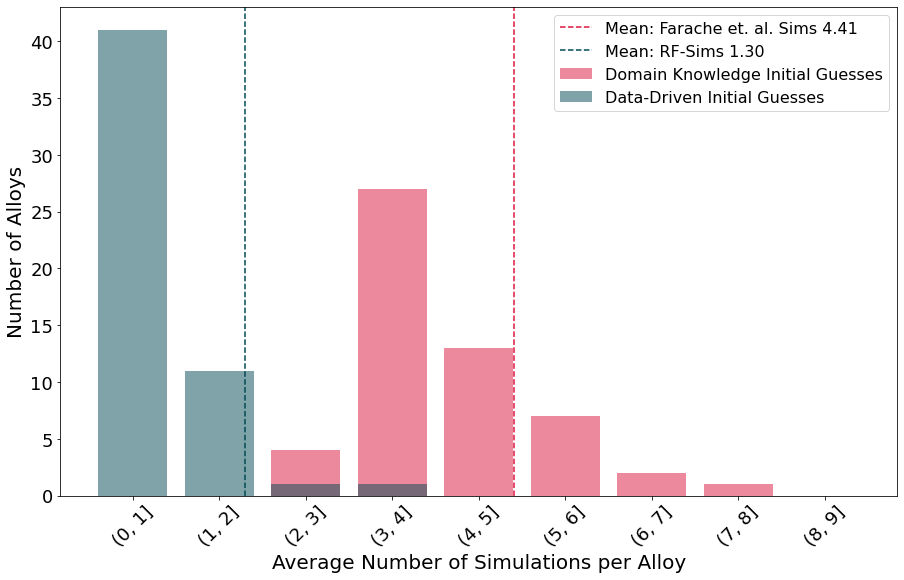}
    \caption{Comparison of average number of simulations per alloy composition with and without a data-driven model for $T_{liq}$ and $T_{sol}$ temperatures.}
    \label{simulation_counts}
\end{figure*}

Using the sequestered test set from the RF training, 54 compositions, we analyzed the number of simulations required to achieve convergence and coexistence. The new model derived from the FAIR data in the ResultsDB clearly outperforms original approach. Figure \ref{simulation_counts} shows that our model reduces the number of simulations per composition from an average of 4.4 to 1.3.

\subsection{AL search of the alloy with the lowest melting temperature}
\label{activelearning_min_section}

To further demonstrate the efficiency improvements of our approach and available data, we used an AL workflow to find alloys with lowest melting temperature, as opposed to the search for the alloy with highest melting temperature explored in \textit{Work 1}. The FAIR data in the ResultsDB enabled fine-tuning the simulations and the development of an accurate RF model to start the iterative optimization resulting in significant decrease in the number of simulations needed to find an alloy with optimal melting temperature. To ensure a fair comparison, we modified our random forest to match the original implementation, using 350 estimators. Using the same acquisition functions modified to perform the reverse optimization (minimization), most acquisition functions were able to identify the alloy with the minimum melting temperature within 3 iterations of different alloys, and $\sim$6 simulations (approximately 2 simulations per alloy to determine its melting temperature). 

\begin{figure}[H]
    \centering
    \includegraphics[scale = 0.5]{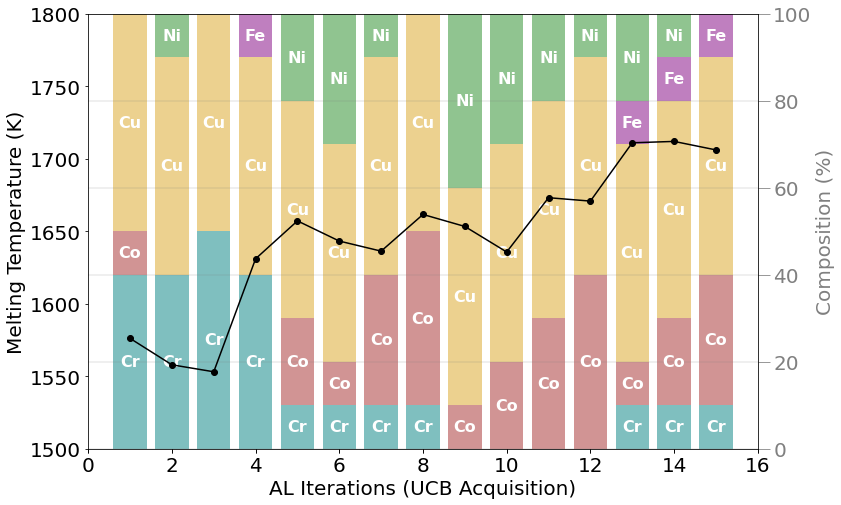}
    \caption{Active learning run using the UCB activation function}
    \label{ucb_activation}
\end{figure}

Figure \ref{ucb_activation} shows the AL iterations performed with the upper confidence bound (UCB) acquisition function. Figures S\ref{figsup:mm_activation} - S\ref{figsup:mei_activation}, describing the remaining acquisition functions, are included in the Supplemental information. We observed a high degree of similarity between alloys chosen by various acquisition functions. In addition to each acquisition function selecting 13 of the same alloys (out of 15 iterations), the order was remarkably similar. All alloys selected (across all acquisition functions) contained exactly 50\% copper. Additionally, we noted that the acquisition functions tended to choose alloys very similar to each other in subsequent iterations, often only varying by 10\% compared to the previous alloy.

Melting temperatures for the same alloy compositions were computed multiple times independently for each acquisition function, resulting in small deviations of approximately 10 K. Alloy compositions found to represent the minimum melting temperature were found within the budget for the acquisition functions, and are as follows: Cr 40\%, Cu 50\%, Ni 10\% (1566 ± 11 K); Cr 40\%, Cu 50\%, Co 10\% (1564 ± 9 K); and Cr 50\%, Cu 50\% (1535 ± 12 K). Uncertainty reported is the standard deviation between independent runs. Notably, the 50\% Cu compositions we identified show melting temperatures consistent with the low values predicted for Cu-rich systems in \textit{Work 1}, validating our AL results. 

Comparing our melting temperature minimization problem to the original maximization effort in \textit{Work 1} could be misleading if the two problems had different complexity. To address this, we also applied the original workflow (\textit{Work 1}) to find the alloy with the lowest melting temperature using the original dataset of 40 alloys, stored in nanoHUB's ResultsDB, and preserved the same search space. The original task, in average, exploring 15 compositions and approximately 60 simulations (about 4 simulations per alloy) to identify the alloy with the highest melting temperature from an unknown set of 555 compositions. The minimization task using the original training set and workflow required 16 compositions (65 simulations) using the UCB acquisition function. Using the MLI acquisition function, the alloy was not found within 40 compositions. 

Overall, we demonstrate a significant increase in search efficiency in terms of number of compositions explored and number of simulations per composition. Across all acquisition functions tested, we demonstrate a 10x increase in performance compared to the results showcased in \textit{Work 1}.

\section{Conclusions}
\label{conclusions_section}

In this work, we demonstrated the transformative potential of FAIR (Findable, Accessible, Interoperable, and Reusable) data and workflows in computational materials science building on modern infrastructure. Building on a published tool for online MD simulations and prior results automatically captured we showed a 10-fold reduction in the resources required for the optimization of a material property as compared to our prior work, \textit{Work 1} \cite{farache2022active}. The latter required characterizing approximately 15 multi-principal component alloys to find the composition with the highest melting temperature. Each composition required, in average, 4 simulations to find the melting temperature. We learned from the FAIR data collected during this initial effort to  i) reduce the number of physics-based simulations required to determine the melting temperature of an alloy given a composition and ii) accelerate exploration and discovery of new alloys using active learning by starting with a more accurate model. Armed with a more accurate model and better starting parameters for the physics-based simulations we are able to find the alloy with the lowest melting temperature with only two simulations.

Our work provides a concrete example of how embracing FAIR data principles can accelerate innovation in materials science. We note that the general concepts and ideas apply well beyond the specific example shown here. In this work reuses a single FAIR workflow for a single quantity of interest, we envision future efforts combining data from multiple tools and various properties. Techniques such as transfer learning and multi-task learning could be used to learn from multiple sources and combine multiple tools.

\section*{Acknowledgements}

This effort was supported by the US National Science Foundation FAIROS program, award 2226418. This effort was also supported by the US National Science Foundation (DMREF-1922316). We acknowledge computational resources from nanoHUB and Purdue University through the Network for Computational Nanotechnology. We also thank Jorge Hernandez, Varun Vaidyanathan, and Kathleen O'Sullivan for their helpful discussion and efforts.

\section*{Data availability}

The FAIR workflow described is available as a nanoHUB tool \cite{nh_fairmeltheas}. The simulation workflow to calculate melting temperature (meltheas) \cite{nh_meltheas} is available as a sim2l \cite{hunt2022sim2ls} in nanoHUB. 

\section*{Ethics declarations}

\subsection*{Conflict of interest}

On behalf of all authors, the corresponding author states that there is no conflict of interest.

\subsection*{Author Information}

\textbf{Corresponding Author}

\textbf{Alejandro Strachan} - School of Materials Engineering and Birck Nanotechnology Center, Purdue University, West Lafayette, Indiana 47907; Email: strachan@purdue.edu \\

\bibliographystyle{unsrt}
\bibliography{references.bib}

\renewcommand\listingscaption{Figure S}
\setcounter{listing}{0}
\renewcommand\figurename{Figure S}
\setcounter{figure}{0}

\section*{Supplemental Information}

\subsection*{Information Acquisition Functions}
\label{acquisition_section}

Formulations for the acquisition functions in Farache et al. \cite{farache2022active} were left unchanged; however, in order for them to perform a minimization, we negated the target values for the function.

\begin{equation}
    MM:  x^*= argmax \;\; -E[M(x_i )]
    \label{eq:MM}
\end{equation}

\begin{equation}
    UCB:  x^*= argmax \;\;  -(E[M(x_i )] + K * \sigma[M(x_i )])
    \label{eq:UCB}
\end{equation}

\begin{equation}
    MLI:  x^*= argmax \;\; - (\frac{E[M(x_{i})] - E[M(best)]}{\sigma[M(x_i )]})
    \label{eq:MLI}
\end{equation}

\begin{equation}
    \begin{split}
        MEI: x^*= argmax \;\; - \rho \; (E[M(x_{i})] - E[M(best)], \sigma [M(x_i)] ) \\
        with \;\;
        \rho  \; (z,s) =  \begin{cases} 
          s \phi'(\frac{z}{s}) +
          z \phi(\frac{z}{s}) & s > 0 \\
          max(z,0) & s = 0
        \end{cases}
    \end{split}    
    \label{eq:MEI}
\end{equation}

\begin{figure}[H]
    \centering
    \includegraphics[scale = 0.5]{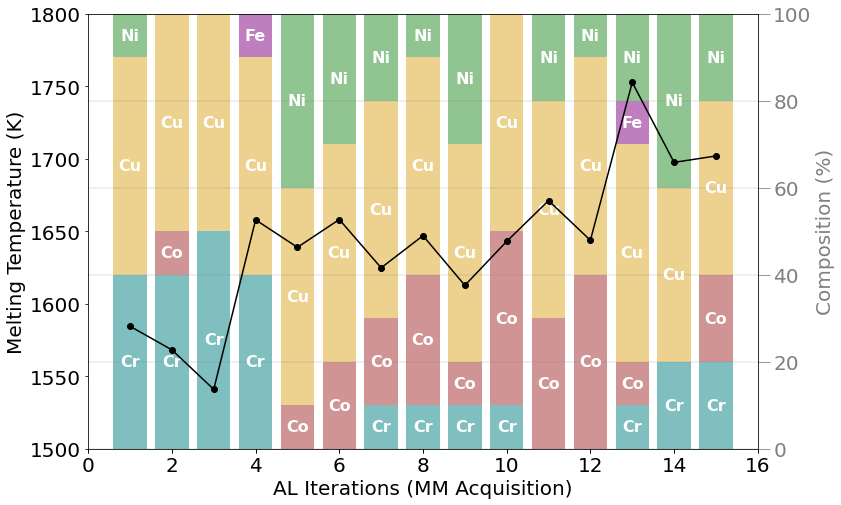}
    \caption{Active learning run using the MM activation function}
    \label{figsup:mm_activation}
\end{figure}

\begin{figure}[H]
    \centering
    \includegraphics[scale = 0.5]{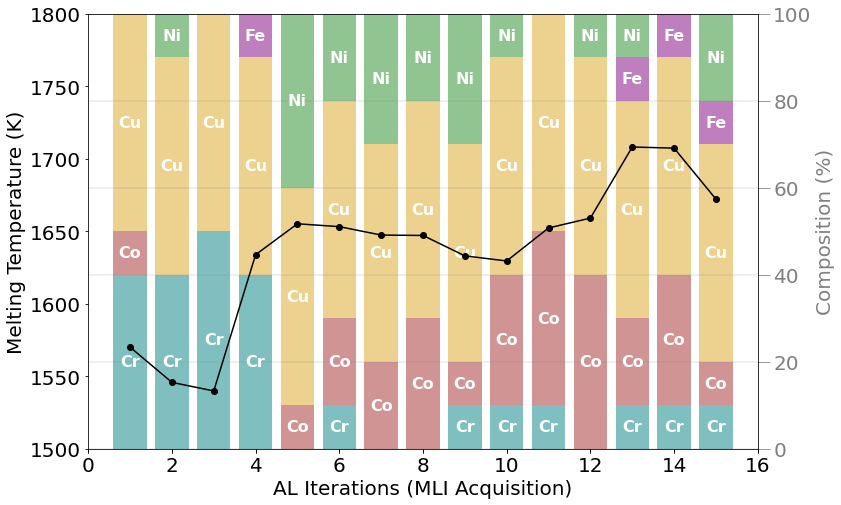}
    \caption{Active learning run using the MLI activation function}
    \label{figsup:mli_activation}
\end{figure}

\begin{figure}[H]
    \centering
    \includegraphics[scale = 0.5]{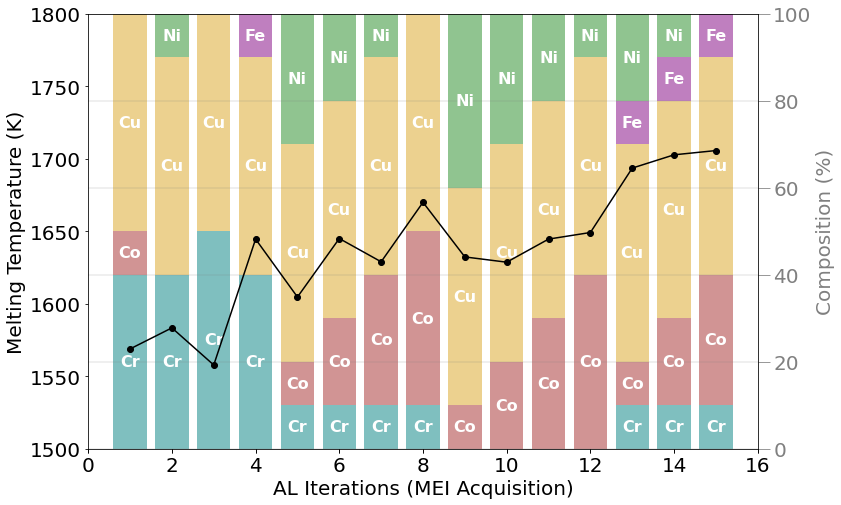}
    \caption{Active learning run using the MEI activation function}
    \label{figsup:mei_activation}
\end{figure}

\begin{figure}[H]
    \centering
    \includegraphics[scale = 0.4]{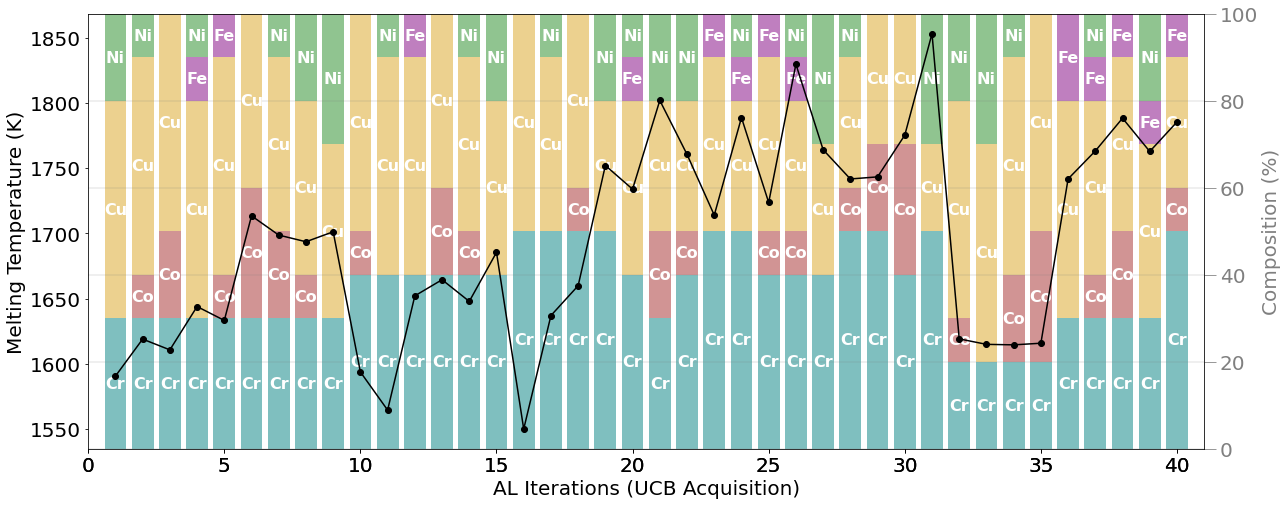}
    \caption{Minimization of MPCA melting temperature using UCB and Farache et. al.'s active learning workflow}
    \label{figsup:farache_min_ucb}
\end{figure}

\begin{figure}[H]
    \centering
    \includegraphics[scale = 0.4]{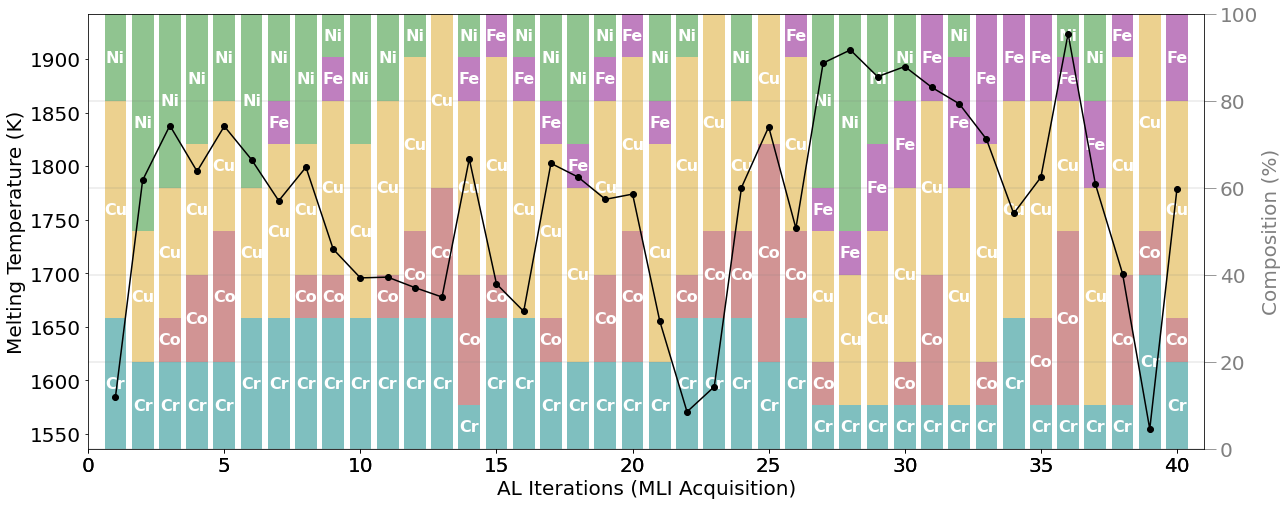}
    \caption{Minimization of MPCA melting temperature using MLI and Farache et. al.'s active learning workflow}
    \label{figsup:farache_min_mli}
\end{figure}

\end{document}